\documentclass[twocolumn,aps,prl,preprintnumbers,amsmath,amssymb,nofootinbib]{revtex4-1}\usepackage[dvips]{graphicx}
\usepackage{color}
\usepackage{ulem}
\begin{document}

\title{Block Copolymer Films with Free Interfaces: Ordering by Nano-Patterned Substrates}

\author{Xingkun Man, David Andelman$^{*}$}
\affiliation{Raymond and Beverly Sackler School of Physics and Astronomy,
Tel Aviv University, Ramat Aviv 69978, Tel Aviv, Israel}

\author{Henri Orland}
\affiliation{Institut de Physique Th\'eorique, CE-Saclay, CEA,
F-91191 Gif-sur-Yvette Cedex, France}


\begin{abstract}

We study  block copolymers (BCP) on patterned  substrates, where the top polymer film surface is not constrained but  is free and can adapt its shape self-consistently. In particular, we investigate the combined effect of the free interface undulations with the wetting of the BCP film as induced by nano-patterned substrates.
In wetting conditions and
for a finite volume of BCP material, we find  equilibrium droplets composed of coexisting perpendicular and parallel lamellar domains.
The self-assembly of BCP on topographic patterned substrates was also investigated and it was found that the free interface induces mixed morphologies of parallel and perpendicular domains coupled with a non-flat free interface.
Our study has some interesting consequences for experimental setups of graphoepitaxy and nanoimprint lithography.

\end{abstract}

\maketitle

Thin films of block copolymer (BCP) have attracted abiding interest over the years due to their numerous industrial applications, such as templates and scaffolds for the fabrication of nanostructured materials where many of the essential attributes of photolithography including pattern perfection, registration, and arbitrarily shaped geometries can be reproduced on the nano scale. As such, it has been considered as a legitimate next-generation microelectronics lithography technique for insertion at the sub-22 nm technology nodes \cite{Li07,ITRS09}.

When thin films of BCP are produced, they self-assemble into ordered nanostructures with a characteristic length scale in the 10-100 nm range. Such self-assembled structures
mainly depend on chain architecture, film thickness and surface interactions. In many theoretical studies~\cite{Matsen97,Petera98,Tsori01} the BCP films are taken to be bound between two flat solid surfaces, and the film structure is influenced by surface interactions and the constrained film thickness. However, in experiments~\cite{Coulon89,Knoll02,Knoll04} the BCP film is spin casted from solution, and after solvent evaporation, the film is bound only by the bottom substrate while the top is a  free interface with air (vapor). Because the film can locally adjust its  height, new phases, which cannot be found when the film is confined between two planar surfaces, can emerge. For example, terrace formation has been found in lamellar~\cite{Coulon89} or cylindrical BCP thin films~\cite{Knoll02,Knoll04}.

In recent experiments~\cite{Stoykovich05,Ruiz08,Bang09,Segalman01,Stein07,Li04,Voet11} employing chemically and topographically patterned surfaces with preferential local wetting properties toward one of the two polymer blocks resulted in unique organization of BCP thin films, as inspired by the demands of the microelectronic industry. In order to have an easily implemented and low cost technique, the typical length scale of the patterns has to be much larger than the BCP natural periodicity, $l_0$. However, such a sparse arrangement of stripes results in additional complications as the free interface starts to undulate and may even change the film morphology in ways that are only poorly understood at present.

The aim of this Letter is to study the complex and interesting interplay between the BCP/air free interface and  nano-patterned substrates. More specifically,
we address theoretically this coupling effect on the self-assembly of  BCP films using self-consistent field theory (SCFT), and find a strong correlation between the ordering induced by the substrate and the interface profile.
Depending on the patterned substrate, we find an undulating free interface or droplet formation, even in {\it complete wetting} conditions.

In most calculations~\cite{Matsen97,Petera98,Tsori01} the interface is modeled by a fixed boundary condition. In these cases, such an adjustment of the thickness of the BCP layer with respect to the BCP ordering cannot take place, resulting in a very different ordering.
To circumvent this difficulty,
we model the BCP film using  SCFT  adapted for the free interface case~\cite{Muller00,Sharma00,Morita01,Lyakhova06,Matsen06}.
The system is composed of polymer chains and a bad solvent (air) inside a fixed volume $V$. The average BCP volume fraction is ${\Phi}_P=\Phi_A+\Phi_B$ where $\Phi_A$ and $\Phi_B$ are the  average volume fractions of the two blocks.
In order to model a strong phase separation
between the polymer and its vapor (air), we add 2nd and 3rd  virial terms to the system free energy:
$-\frac{1}{2}v\phi_P^2({\bf r})+\frac{1}{3}w\phi_P^3({\bf r})$, where $\phi_P({\bf r})$ is the local volume fraction of the polymer. For $v>0$ (bad solvent),
the system will undergo demixing and a 3rd virial term with $w>0$ is needed as a stabilizing term.
The two coexisting phases are a polymer-rich phase with $\Phi_P^*\approx v/w$ and  a vapor-rich phase, $\Phi_P^* \ll 1$.
In all calculations hereafter, we choose $v=2.6$ and $w=1.8$.
The resulting two-dimensional free energy is then:

\begin{eqnarray}\label{f1}
\frac{b^2 F}{k_BT}=\int d^2{\bf r}\left[\chi_{_{\rm AB}}\phi_{A}({\bf r})\phi_{B}({\bf r})
-\frac{{\Phi}_P}{N}\ln Q_c \right.\nonumber\\
-\,\omega_A({\bf r})\phi_A({\bf r})-
 \omega_B({\bf r})\phi_B({\bf r})\nonumber\\
\left. -\, \frac{v}{2}(\phi_A({\bf r})+\phi_B({\bf r}))^2+\frac{w}{3}(\phi_A({\bf r})+\phi_B({\bf r}))^3 \right]\nonumber\\
-\,b\int dx\left[\Delta u_A\phi_A(x,0)+\Delta u_B\phi_B(x,0)\right]
\end{eqnarray}
The Flory-Huggins parameter $\chi_{_{\rm AB}}$ represents the interaction between the A and B blocks, $\omega_{j}({\bf r})$, $j={\rm A, B}$, are the two auxiliary fields coupled with $\phi_{j}({\bf r})$, and $Q_{c}$ is the single-chain partition function in the presence of the $\omega$-fields. The two surface parameters are $\Delta u_{A}=u_A-u_S$ and $\Delta u_{B}=u_B-u_S$, where $u_A$, $u_B$ and $u_S$ represent the short-range interaction of the substrate with the A block, the B block and vapor, respectively. All lengths are expressed in terms of $b$, the Kuhn length,   $k_B$ is the Boltzmann constant, and $T$ is the temperature. For simplicity, each of the BCP chains consists of $N=N_{A}+N_{B}$ monomers and is symmetric with $N_{A}=N_{B}$.

Using the saddle-point approximation, we obtain a set of self-consistent equations
that are solved numerically. As is known from previous works~\cite{Leibler80,Matsen94,Man10}, symmetric BCP in the bulk have a transition between a disordered phase above the Order-Disorder temperature (ODT), $N\chi_c\simeq 10.5$, and a lamellar phase of natural periodicity $l_0$, below the ODT.
The numerical procedure, i.e. starting with an initial profile and iterating until convergence, allows us to determine self-consistently the morphological changes inside the BCP film as well as the structure of the non-flat BCP/air interface. Note that due to the existence of many metastable states, the converged state may strongly depend on the initial conditions.

\begin{figure}[h]
\begin{center}
{\includegraphics[bb=0 0 373 285, scale=0.5,draft=false]{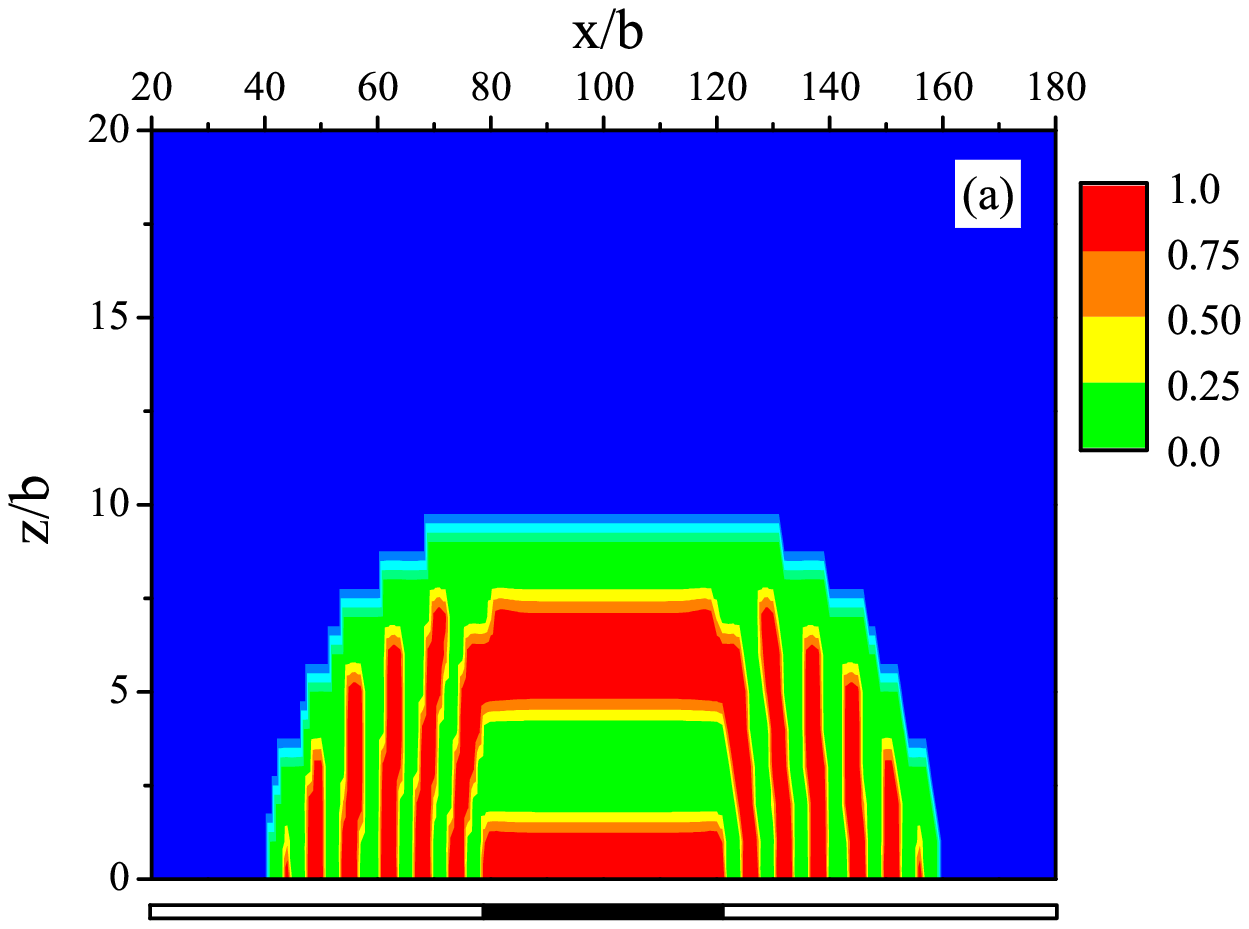}}
{\includegraphics[bb=-30 0 373 285, scale=0.5,draft=false]{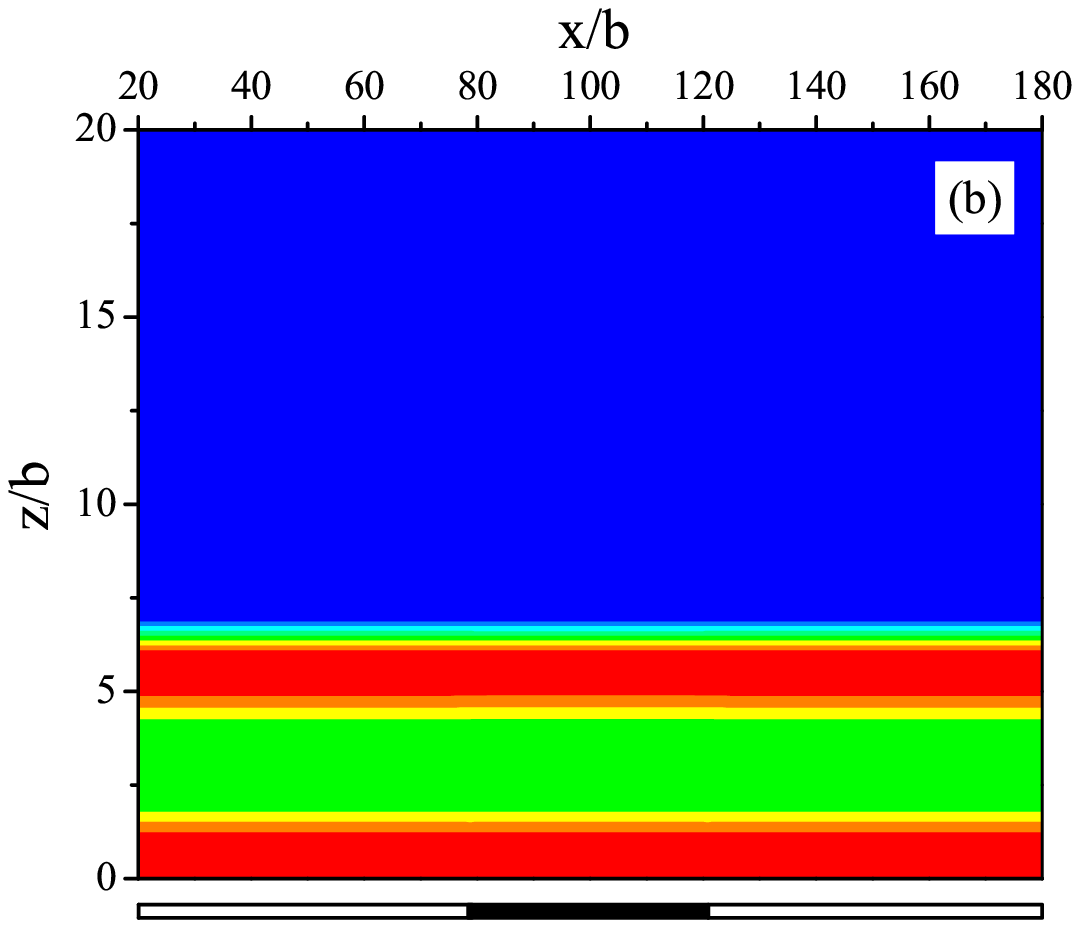}}
\caption{\textsf{(color online). Contour plots of $\phi_A$ in the lamellar BCP phase with $N\chi_{_{\rm AB}}=26$ in contact with a chemically patterned surface. For the middle surface stripe (black) $\Delta u_A=0.6$, $\Delta u_B=0.4$, while for the rest of the substrate (white) $\Delta u_A=\Delta u_B=0.4$. The size of the calculation box is $L_x\times L_z=180b\times20b$ and $l_0\simeq 7b$. (a) ${\Phi}_P=0.1$ and the average film thickness is $h_0\approx0.6l_0$.
 (b) ${\Phi}_P=0.175$ and the average film thickness $h_0\approx l_0$. The values of the color code vary between 0 and 1. The air is  defined operationally in regions where the total polymer density is less than 0.05 and is represented by the dark blue (black) color. Only part of the calculation box is shown, and note the different scale of the $x$ and $z$ axes.}}  \label{fig1}
\end{center}
\end{figure}

\begin{figure}[h]
\begin{center}
{\includegraphics[bb=0 0 373 285, scale=0.5,draft=false]{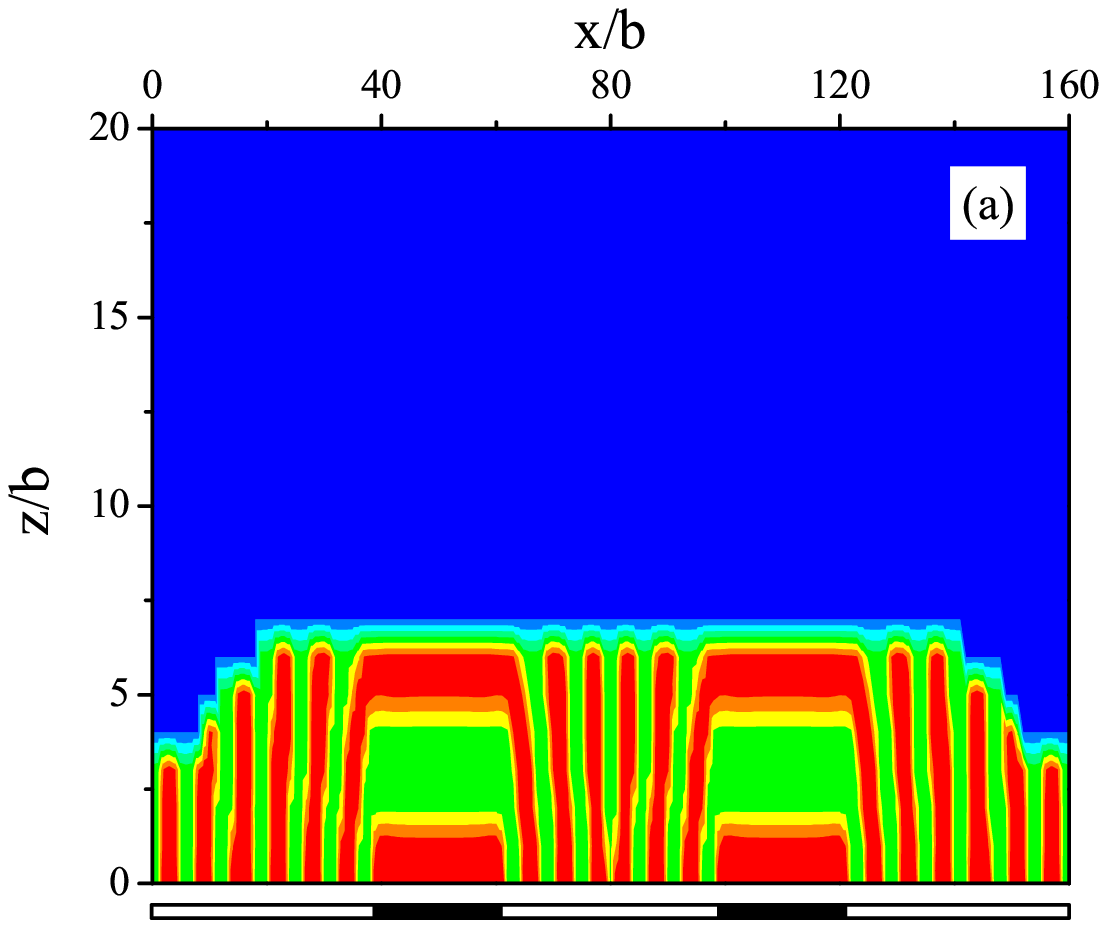}}
{\includegraphics[bb=0 0 373 285, scale=0.5,draft=false]{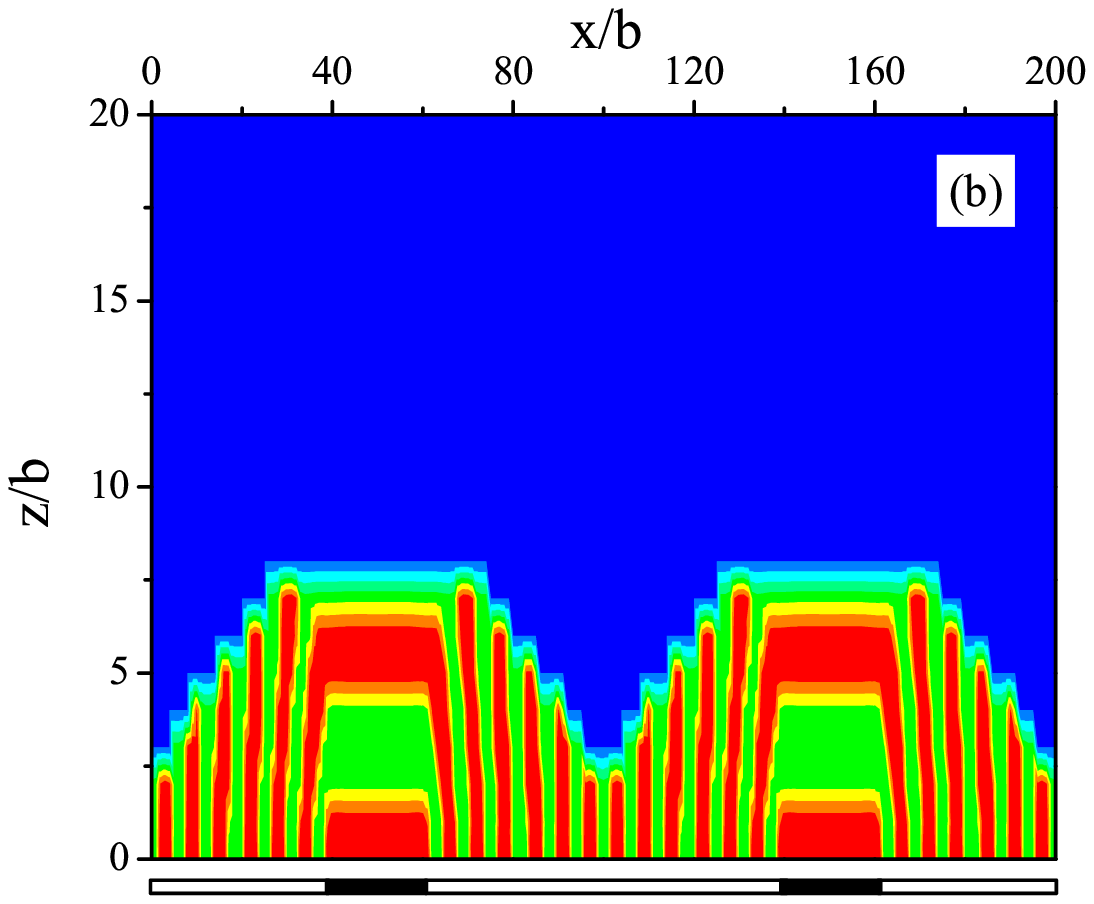}}
\caption{\textsf{(color online). Contour plots of $\phi_A$ in the lamellar BCP phase with fixed  ${\Phi}_P=0.2$, $\l_0\simeq 7b$ and $N\chi_{_{\rm AB}}=26$ (below ODT). (a) Two chemical surface stripes (black) have a width of $l_s=20b$ with $\Delta u_A=0.6$ and $\Delta u_B=0.4$, while the rest of the surface (white) is neutral with $\Delta u_A=\Delta u_B=0.4$. The inter-stripe region has a width of $l_n=40b$. (b) Same film as in (a) but with a larger inter-stripe distance of $\l_n/b=80$. The size of the calculation box is $L_x\times L_z=160b\times40b$ in (a) and $L_x\times L_z=200b\times40b$ in (b). The color code and height of the plotted box are the same as in Fig.~1.}}
\label{fig2}
\end{center}
\end{figure}

\begin{figure}[h]
\begin{center}
{\includegraphics[bb=0 0 373 285, scale=0.5,draft=false]{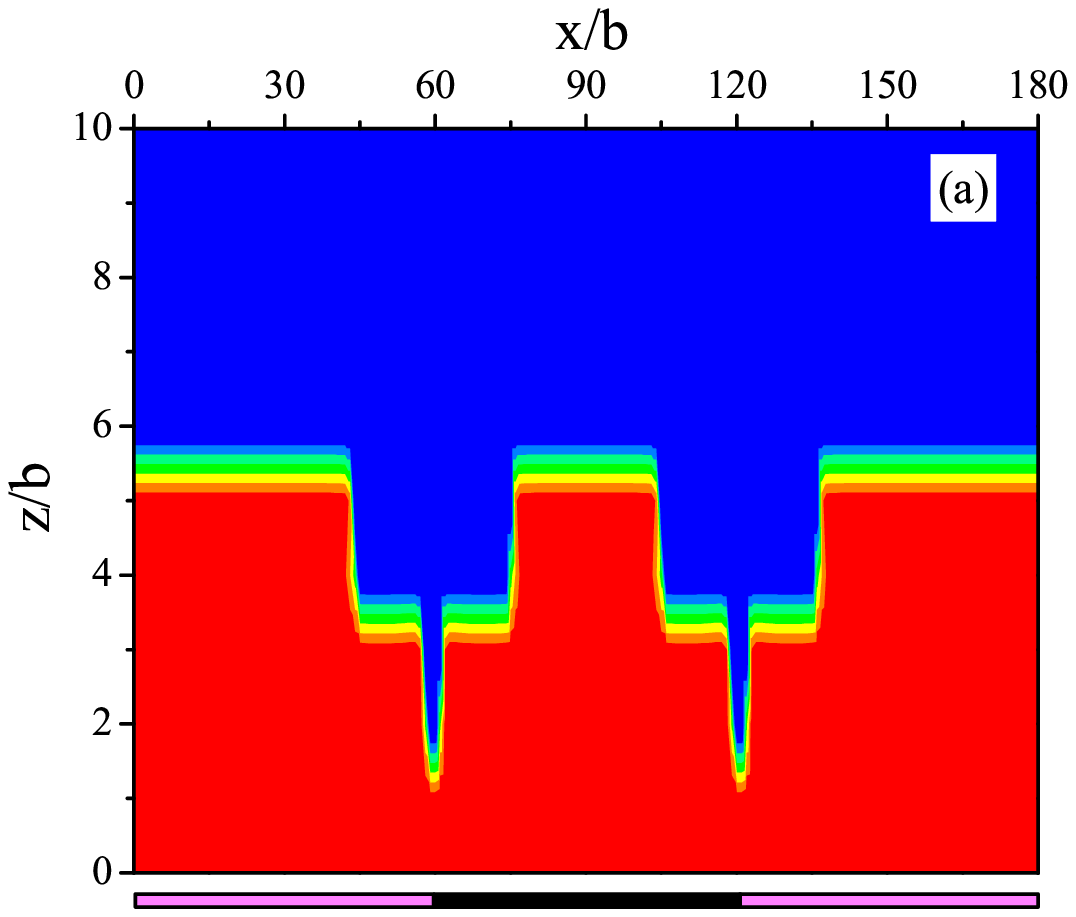}}
{\includegraphics[bb=0 0 373 285, scale=0.5,draft=false]{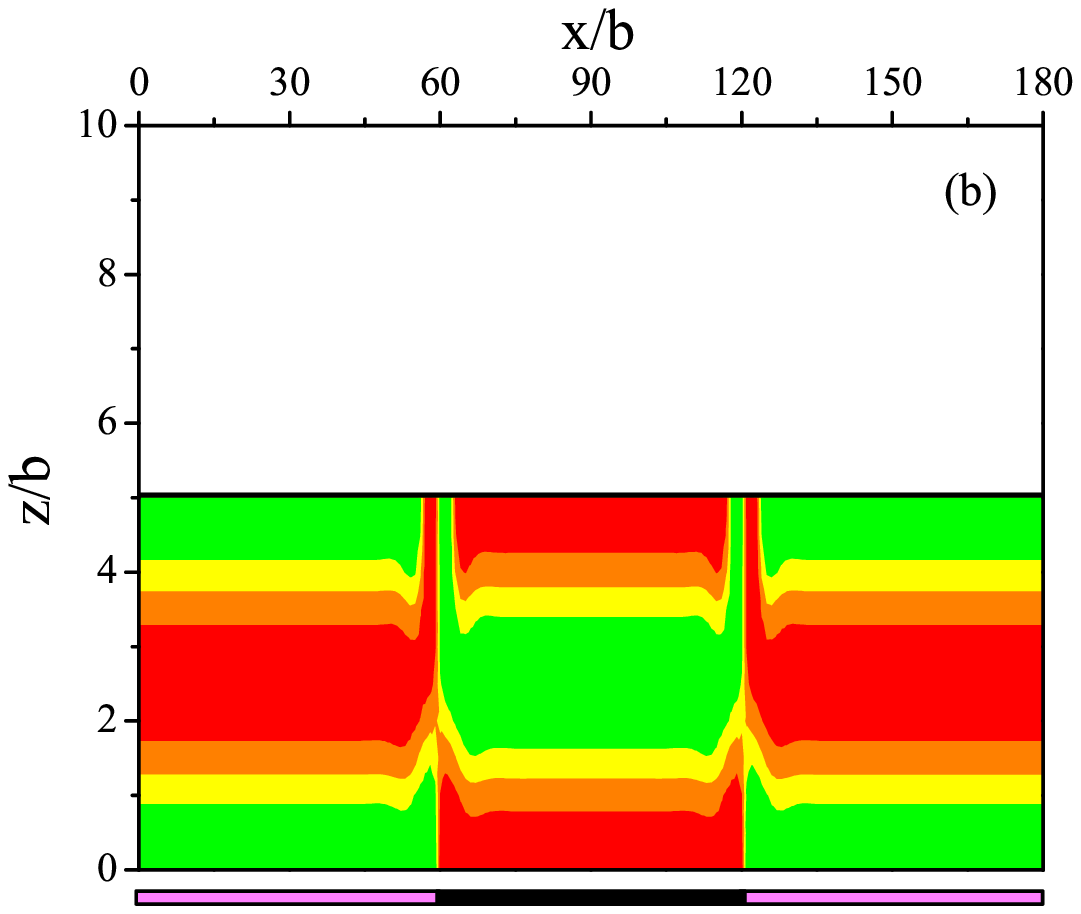}}
\caption{\textsf{(color online). Lamellar BCP structures with fixed total BCP volume fraction ${\Phi}_P=0.12$, $\l_0\simeq 7b$ and with $N\chi_{_{\rm AB}}=26$ (below ODT). In (a) the polymer profile $\phi_P$ is shown. Two chemical stripes  of width $l_s=60b$ and $u_A=0.4$ and $u_B=0.6$
are separated by a central stripe (black) of width $l_s=60b$ and $u_A=0.6$ and $u_B=0.4$. For comparison, in (b) the A-block profile $\phi_A$ is shown for a planar BCP film bound by two solid and flat surfaces at $z=0$ and $z=5b$. The patterned substrate and all other parameters are the same as in (a). The size of the calculation box  is $L_x\times L_z=180b\times40b$ in (a) and  $L_x\times L_z=180b\times5b$ in (b). The color code is the same as in Fig.~1.}}
\label{fig3}
\end{center}
\end{figure}

We first present  some results on the adsorption of BCP films on chemically patterned substrates in complete wetting conditions.
The BCP film is spread on top of a planar solid substrate modeled as a series of chemical stripes of width $l_s$ and inter-stripe separation $l_n$. All the stripes have a constant preference for the A block, e.g., $\Delta u_A=0.6$, $\Delta u_B=0.4$, while the inter-stripe regions are taken to be neutral, e.g., $\Delta u_A=\Delta u_B=0.4$.  It can easily be seen that this corresponds to complete wetting conditions for both blocks as seen in Fig.~\ref{fig1}(b). The figure shows that for our chosen system parameters, the BCP films completely wets the substrate in the case when the film thickness is commensurate with the natural period of the BCP, $l_0$.

For the case of a single stripe in the center of a large substrate,  although we are in wetting conditions, a BCP droplet can be obtained (Fig.~\ref{fig1}a). We choose an amount of BCP  such that its homogeneous spreading over the entire substrate is equal to a non-integer number of parallel lamellae, $h_0=0.6 l_0$.
Starting from a localized ``rectangular"-shaped lamellar BCP droplet as an initial condition,  after convergence we obtain
a BCP droplet residing on the substrate as shown in Fig.~\ref{fig1}a. It is composed of coexisting perpendicular and parallel lamellae in registry with the patterned substrate.  The free-energy density of this droplet is $f=-0.0535$.
In contrast, when the initial condition is taken to be a flat BCP film above its ODT,
 a parallel lamellar phase is obtained with free energy density $f=-0.0520$, clearly larger than that of the droplet.
The droplet shown in Fig.~1a is more stable because the parallel domain height is larger than $h_0$ as it tries to accommodate an integer number of layers.
Consequently, as the BCP volume is fixed, the height of these perpendicular domains should compensate that of the
parallel ones.

In certain cases the lamellar phase can be the more stable one (Fig.~1b). Performing the same set of calculations as above with an amount of BCP corresponding to one lamellar period, $h_0=l_0$,
we find that the lamellar phase has a free energy density $f=-0.0998$, and is more stable than that of the droplet with $f=-0.0957$.

It is natural to generalize the results of Fig.~1 to more than one surface stripe. An undulating BCP/air interface in this case follows in registry the stripe structure of the surface. This is shown in Fig.~2, where  the substrate has two chemical stripes.  Because different surface regions induce parallel and perpendicular lamellae structures, their coexistence yields an undulating BCP/air interface (Fig.~2b). Moreover, we find that the undulating free interface  depends on the inter-stripe distance $l_n$. If $l_n$ becomes small~\cite{numerical}, the BCP/air interface between the two stripes is almost flat (Fig.~2a). This means that the stripes cannot be too sparse (as desired in many applications) if one wants to obtain a flat free interface.

Further difference between the free interface and the flat (and rigid) interface can be seen in Fig.~3. The patterned surface is chosen with alternating stripes preferring A and B layers inducing parallel lamellar domains. As can be seen in Fig.~3a, the overall polymer profile, $\phi_P({\bf r})$ displays several terraces at the free interface. They are induced  by  defects in the film (not seen in Fig.~3a), which are themselves created because of the patterned substrate. The terraces are in registry with the substrate stripe boundaries and the jumps in the terrace height correspond to half of the lamellar periodicity. The situation is rather different in Fig.~3b where the top interface is forced to be planar. The film has the same total volume as in 3a, yielding a constant height $h_0=5b$. However, differences in the A-block profile, $\phi_A$, as induced by the surface stripes preferring the A or B lamellae are clearly seen.

We also considered the situation where BCP films are in contact with topographic patterned substrates used in graphoepitaxial~\cite{Segalman01,Stein07,Sharma08} and nanoimprint lithography~\cite{Li04,Voet11,Man11,Thebault12} set-ups.
The surface is modeled as a sequence of grooves that have a slight preference toward one of the blocks (say, the A block).
When the  BCP volume exceeds the inner volume of the groove, we obtain a mixed state composed of parallel and perpendicular lamellar domains (Fig.~4a). One of our conclusions is that perfect perpendicular lamellae structure can be obtained when the BCP is bounded by a top flat solid surface, as can be seen in Fig.~4b. This suggests that pressing the film between the grooved mold and a second solid surface facilitates the formation of a perpendicular lamellar orientation
and is superior to the free interface set-up, as in Fig.~4a.

\begin{figure}[h]
\begin{center}
{\includegraphics[bb=0 0 337 276, scale=0.5,draft=false]{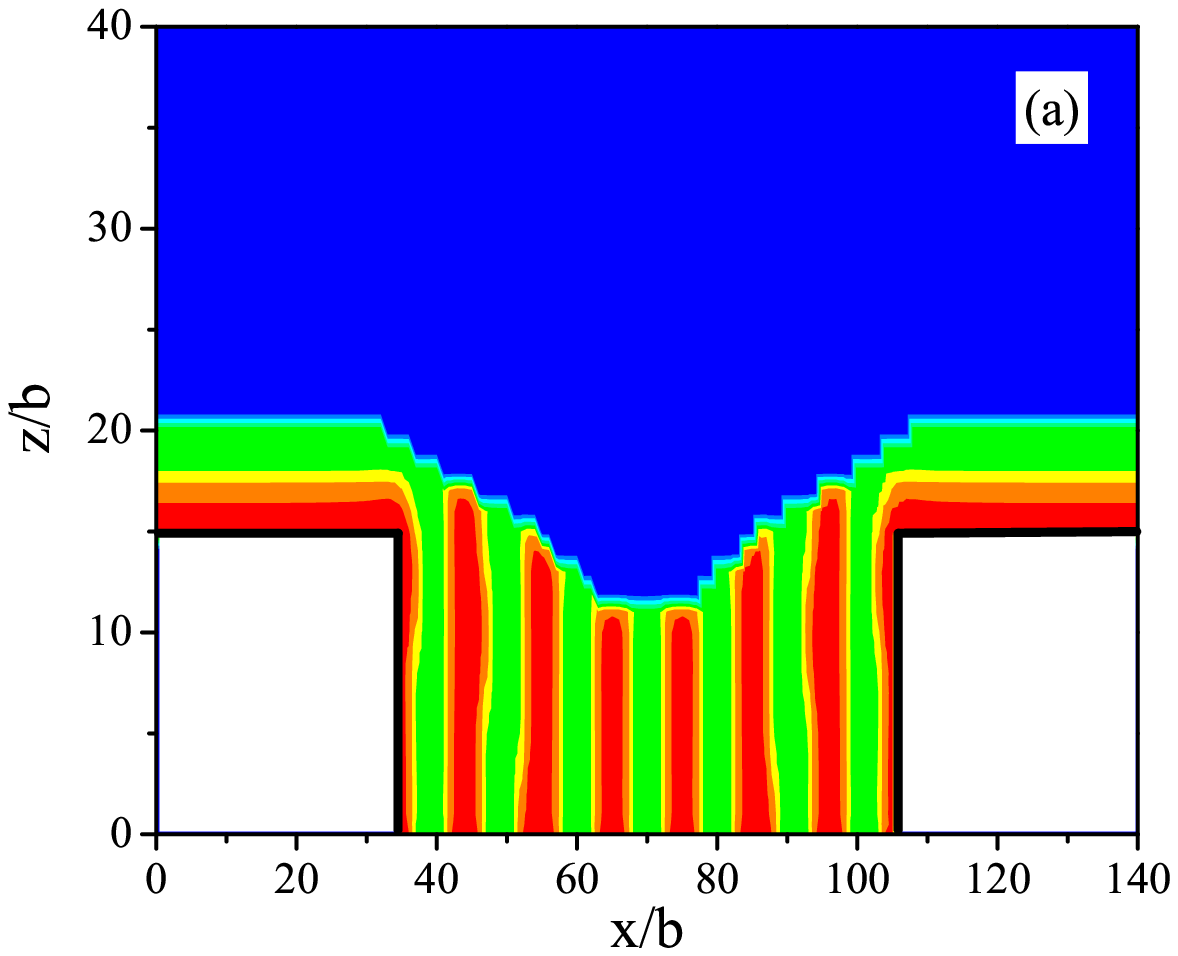}}
{\includegraphics[bb=0 0 337 276, scale=0.5,draft=false]{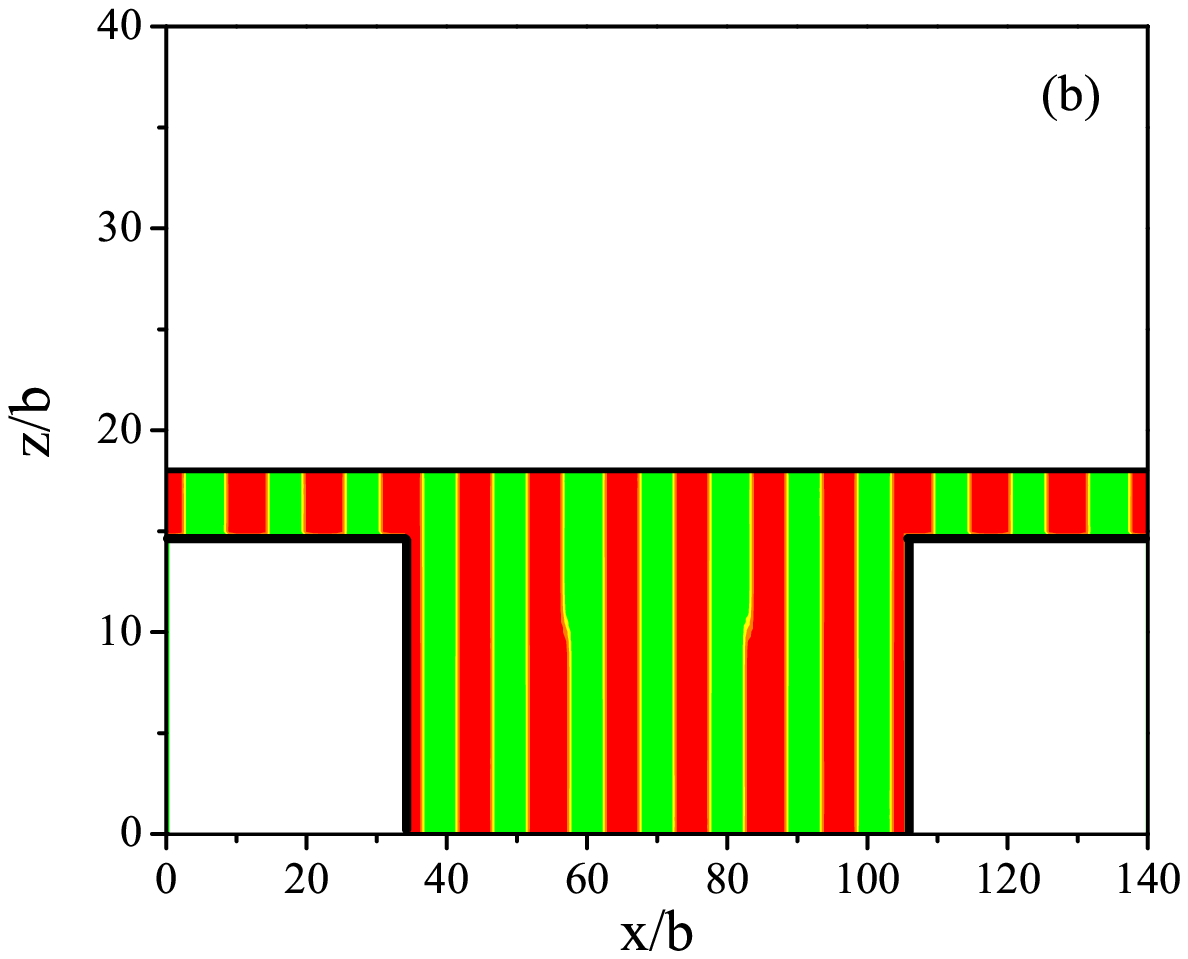}}
\caption{\textsf{(color online). BCP lamellae structures for topographical surfaces (grooves) with ${\Phi}_P=0.38$, $\l_0\simeq 10b$ and $N\chi_{_{\rm AB}}=43.2$ (a) Contour plot of the A-block, $\phi_A$. $\Delta u_A=0.12$ and $\Delta u_B=0.1$ for the vertical and side walls, while $\Delta u_A=\Delta u_B=0.1$ for the bottom of the groove. (b) The BCP is bound from above by a second solid surface, leading to a perfect perpendicular orientation. All other parameters are the same as in (a). The color code is the same as in Fig.~1.}} \label{fig4}
\end{center}
\end{figure}

In this Letter we  addressed the combined effects of  nano-patterned substrates  and the polymer/air free interface on the self-assembly of BCP thin films. We treated air as a bad solvent and modified accordingly the SCFT in order to study the phase behavior of finite volume films.
We obtained an undulating polymer/air interface when the films are casted on chemically patterned substrates. For a sequence of chemical stripes, we find that the undulations of the BCP/air interface  depends on the inter-stripe spacing. In particular, by adjusting the substrate affinity, we can obtain BCP droplets containing mixed perpendicular and parallel lamellae. We also studied the self-assembly of BCP thin films on topographic substrates. If the BCP amounts exceeds the groove inner volume, we obtain a mixed perpendicular and parallel lamellae structure instead of  a perfect 2D perpendicular lamellae. Furthermore, we find that it is more efficient to confine a BCP film between a solid mold and a counter planar surface in order to obtain a perfect perpendicular lamellar structure.

Our SCFT calculations show new and interesting phases of BCP thin films where the free interface displays height variations coupled with morphological changes as determined by the patterned substrate. Even small height variations allow the relaxation of the frustration due to the incompressibility of the BCP lamellae in contrast to fixed boundary models, leading to BCP ordering that is very different in these two approaches.  We finally note that the polymer-vapor interaction parameter is chosen as $\nu=2.6$, while the AB interfacial tension varies between  $\chi_{\rm AB}=1.3$ (Figs.~\ref{fig1}-\ref{fig3}) and 0.8 (Fig.~\ref{fig4}). This means that the magnitude of $\chi_{\rm AB}$  is not much smaller than  $\nu$ as is the case  for typical BCP systems. For example, for PS (PMMA) used in Ref.~\cite{Matsen06} the surface tension with the air is 31.2\,dyne/cm (31.4\,dyne/cm), while the PS-PMMA interfacial tension is about 1.5\,dyne/cm (at the proper temperature range). Unfortunately, such an extreme tension ratio, $\chi_{\rm AB}\ll \nu$, is troublesome for the SCFT calculation, because much finer grids are required at larger $\nu$  values. The same numerical difficulty has also been reported in Ref.~\cite{Matsen06}.

We hope that in the future, more numerically-intensive calculations for free interfaces will shed additional light on the phase behavior of BCP thin films as influenced by their nano-patterning substrates  and free interfaces.


{\bf Acknowledgements.~~~} We thank D. Ben-Yaakov for useful discussions and suggestions. This work
was supported in part by the U.S.-Israel Binational Science Foundation under Grant No.
2006/055, the Israel Science Foundation under Grant No. 231/08, the Center for Nanoscience and Nanotechnology at Tel Aviv University. One of us (HO) would like to thank  the Raymond \& Beverly Sackler Program for Senior Professors by Special Appointment at Tel Aviv University.



\begin{thebibliography}{99}

\bibitem{Li07} W.-K. Li, and S. Yang, J. Vac. Sci. Technol. B, {\bf25}, 1982 (2007)

\bibitem{ITRS09} {\it International Technology Roadmap for Semiconductors (ITRS) 2009 Edition}, Chapter {\it Lithography}. See http://www.itrs.net.

\bibitem{Matsen97} M. W. Matsen, J. Chem. Phys. {\bf106}, 7781 (1997).

\bibitem{Petera98} D. Petera, and M. Muthukumar, J. Chem. Phys. {\bf109}, 5101 (1998).

\bibitem{Tsori01} Y. Tsori, and D. Andelman, J. Chem. Phys. {\bf115}, 1970 (2001).

\bibitem{Coulon89} G. Coulon, T. P. Russell, and V. R. Deline, Macormolecules, {\bf22}, 2581 (1989).

\bibitem{Knoll02} A. Knoll, A. Horvat, K. S. Lyakhova, G. Krausch, G. J. A. Sevink, A. V. Zvelindovsky, and R. Magerle, Phys. Rev. Lett. {\bf89}, 035501 (2002).

\bibitem{Knoll04} A. Knoll, R. Magerle, and G. Krausch, J. Chem. Phys. {\bf120}, 1105 (2004).

\bibitem{Stoykovich05} M. Stoykovich, M. M$\ddot{\rm u}$ller, S. Kim, H. Solak, E. Edwards, J. J. de Pablo, and P. F. Nealey, Science {\bf308}, 1442 (2005).

\bibitem{Ruiz08} R. Ruiz, H. M. Kang, F. A. Detcheverry, E. Dobisz, D. S. Kercher, T. R. Albrecht, J. J. de Pablo, and P. F. Nealey, Science {\bf321}, 936 (2008).

\bibitem{Bang09} J. Bang, U. Jeong, D. Y. Ryu, T. P. Russell, and C. J. Hawker, Adv. Mater. {\bf21}, 4769 (2009).

\bibitem{Segalman01} R. Segalman, H. Yokoyama, and E. J. Kramer, Adv. Mater. {\bf13}, 1152 (2001).

\bibitem{Stein07} G. E. Stein, W. B. Lee, G. H. Fredrickson, E. J. Kramer, X. Li, and J. Wang, Macromolecules {\bf40}, 5791 (2007).

\bibitem{Li04} H.-W. Li, and W. T. S. Huck, Nano Lett. {\bf4}, 1633 (2004).

\bibitem{Voet11} V. Voet, T. Pick, S.-M. Park, M. Moritz, A. Hammack, D. Urban, D. Ogletree, D. Olynick, and B. Helm, J. Am. Chem. Soc. {\bf133}, 2812 (2011).

\bibitem{Muller00} M. M$\ddot{\rm u}$ller, and L. G. MacDowell, Macromolecules {\bf33}, 3902 (2000).

\bibitem{Sharma00} K. Kargupta, and  A. Sharma, Phys. Rev. Lett. {\bf 86}, 4536 (2001).

\bibitem{Morita01} H. Morita, T. Kawakatsu, and M. Doi, Macormolecules {\bf34}, 8777 (2001).

\bibitem{Lyakhova06} K. S. Lyakhova, A. Horvat, A. V. Zvelindovsky, and G. J. A. Sevink, Langmuir {\bf22}, 5848 (2006).

\bibitem{Matsen06} U. Kim, and M.~W. Matsen, Soft Matter {\bf 5}, 2889 (2009).

\bibitem{Leibler80} L. Leibler, Macromolecules {\bf 13}, 1602 (1980).

\bibitem{Matsen94} M. W. Matsen, and M. Schick, Phys. Rev. Lett. {\bf 72}, 2660 (1994).

\bibitem{Man10} X. K. Man, D. Andelman, and H. Orland, Macormolecules {\bf43}, 7261 (2010).

\bibitem{numerical}
{Numerically we determine that the critical value of $l_n$ is around $6l_0$, namely, when $l_n\leq 6l_0$ we can only obtain a flat interface between two stripes, as shown in Fig.~2a.}

\bibitem{Sharma08} R. Mukherjee, D. Bandyopadhyay, and A. Sharma, Soft Matter {\bf 4}, 2086 (2008).

\bibitem{Man11} X. K. Man, D. Andelman, H. Orland, P. Th$\acute{e}$bault, P.-H. Liu, P. Guenoun, J. Daillant, and S. Landis, Macromolecules {\bf44}, 2206 (2011).

\bibitem{Thebault12}
 P. Thebault, S. Niedermayer, S. Landis, N. Chaix, P. Guenoun, J. Daillant, X. K. Man, D. Andelman, and H. Orland,
Adv. Mater. {\bf 24}, 1952 (2012).

\end{thebibliography}
\end{document}